\documentclass[aps,pra,twocolumn,superscriptaddress,nofootinbib,floatfix]{revtex4}

\usepackage{amsmath, amssymb}
\usepackage{graphics}
\usepackage{pstricks}
\usepackage{graphicx}
\usepackage{color}

\newcommand{\dm}[4]{\rho_{#1}(#2 #3 #4)}
\newcommand{\op}[1]{\hat #1}
\newcommand{\opdag}[1]{\hat #1^\dag}
\newcommand{\del}{\partial}
\newcommand{\h}{\hbar}

\begin{document}

\title{Delocalization of ultracold atoms in a disordered potential due to light scattering}

\author{Boris Nowak}
\affiliation{Institut f\"ur Theoretische Physik,
             Ruprecht-Karls-Universit\"at Heidelberg,
             Philosophenweg~16,
             69120~Heidelberg, Germany}
\affiliation{ExtreMe Matter Institute EMMI,
             GSI Helmholtzzentrum f\"ur Schwerionenforschung GmbH, 
             Planckstra\ss e~1, 
             64291~Darmstadt, Germany} 

\author{Jami J. Kinnunen}
\affiliation{Department of Applied Physics, Aalto University School of Science, P.O.Box 15100, FI-00076 Aalto, Finland} 

\author{Murray J. Holland}
\affiliation{JILA, NIST and Department of Physics, University of Colorado, Boulder CO 80309-0440, USA}

\author{Peter Schlagheck}
\affiliation{D\'epartement de Physique, Universit\'e de Li\`ege,
all\'ee du 6 Ao\^ut 17, 4000 Li\`ege, Belgium} 

\date{\today}

\begin{abstract}
We numerically study the expansion dynamics of ultracold atoms 
in a one-dimensional disordered potential in the presence of a weak
position measurement of the atoms.
We specifically consider this position measurement  
to be realized by a combination of an external laser and a periodic
array of optical microcavities  along a waveguide. 
The position information is acquired through the scattering of a 
near-resonant laser photon into a specific
eigenmode of one of the cavities.
The time evolution of the atomic density in the presence of this light 
scattering mechanism is described within a Lindblad master equation approach, 
which is numerically implemented using the Monte Carlo wave function technique. 
We find that an arbitrarily weak rate of photon emission
leads to a breakdown of Anderson localization of the atoms.
\end{abstract}

\maketitle

\section{Introduction}
\label{sec:intro} 

The realization of {potentials with} controlled disorder  
for ultracold atoms
by means of optical speckle fields \cite{LyeO05PRL,Sanchez-Palencia2007} 
or bichromatic optical lattices \cite{DamO03PRL} has recently led to the 
observation of Anderson localization with Bose-Einstein condensates 
\cite{Nature1,Nature2}.
In those experiments, atomic Bose-Einstein condensates, prepared in
a harmonic trap, were released into one-dimensional optical waveguides
which were superimposed {with} disordered potentials realized with 
speckle fields \cite{Nature1} as well as with bichromatic optical lattices 
\cite{Nature2}.
Absorption images of the atomic cloud after the expansion process within the
waveguide clearly revealed an exponential decrease of the average atomic 
density with the distance from the center of the former trap, which is the
characteristic signature of Anderson localization \cite{Anderson}.
While interaction effects did not play a role in those pioneering experiments,
more recent studies specifically focus on the interplay of atom-atom 
interaction and localization in disordered potentials 
(e.g.\ Ref.~{\cite{DeiO10NP, Lucioni2011}}).
Current research directions include  the exploration of
Anderson localization with ultracold atoms in three spatial dimensions
\cite{KonO11S}, with the particular aim to study the Anderson 
metal-insulator transition \cite{ChaO08PRL}.

Clearly, a key condition for the observability of Anderson localization
with ultracold atomic gases is the overall coherence of the atomic cloud.
Any mechanism of decoherence would compromise the phenomenon of destructive
wave interference that lies at the heart of Anderson localization 
\cite{Anderson} and thereby give rise to delocalization.
This also concerns any \textit{in-situ} monitoring of the evolution of the 
atomic cloud during its expansion, by intermediate measurements of the 
positions of atoms.
Evidently, the strong refocusing of the atomic wavefunction that results from
a precise position measurement would destroy the coherence of the atom,
enhance its kinetic energy, and eventually let the atom behave as a classical
particle when being performed several times.

\begin{figure}[t]
\centering
\includegraphics[width=\linewidth]{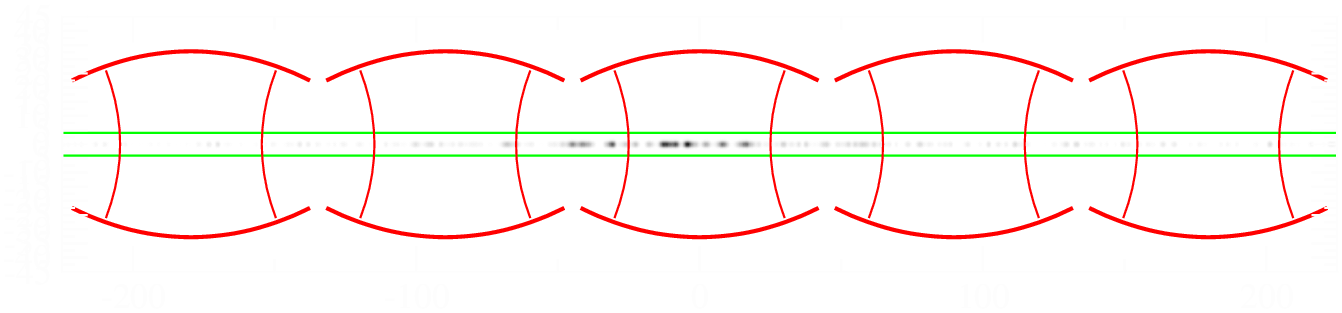}
\caption{\label{setup} (Color online)
Sketch of the configuration under consideration.
An atom, which is initially localized within a harmonic trap at the center,
is expanding within a one-dimensional waveguide (indicated by the green
horizontal lines) to which a disordered potential is superimposed.
A periodic array of optical microcavities (symbolized by the red arcs) is
used to measure the position of the atom, on a length scale that is comparable
to its localization length (indicated by the shadow plot in the waveguide, 
which shows the density of a localized state).
For this purpose, the waveguide is considered to be irradiated by a 
near-resonant laser beam, which may induce spontaneous emissions of photons 
into one of the cavities, possibly to be measured by photodetectors.
The sketch is to scale with the parameters considered in this study, as far
as the horizontal length scales are concerned.
}
\end{figure}

The situation is less obvious for ``weak'' measurement processes, in which
the position of the atom is determined with a large spatial uncertainty
that is of the order of the expected localization length within the disordered
potential.
Such weak measurements might still preserve coherence to a certain extent,
while, at the same time, providing some rough \textit{in-situ} information
on the position of the atom.
One could, for this purpose, conceive \textit{e.g.}\ 
a periodic array of optical 
microcavities placed around the waveguide in which the atoms propagate,
as depicted in Fig.~\ref{setup}.
A near-resonant laser beam which irradiates this configuration can be used
to transfer the atoms to an electronically excited state, from which they
can relax to the ground state by a spontaneous emission of a photon into
one of the cavities, which in turn could be measured by photodetectors 
placed behind the cavities.
The whole configuration could possibly be fabricated on ``atom chips''
\cite{ForZim07RMP}, in which case the disordered potential could arise from
imperfections in the current-carrying wires that generate the magnetic
waveguide potential of the atoms \cite{EstO04PRA, DisBec7}.
Our results below are, however, more general and we expect to see 
the same effects on the localization in the presence of any mechanism of 
similar position measurement.

The aim of this study is to investigate to which extent this approximate
realization of a ``Heisenberg microscope'' gives rise to delocalization 
of an atom in a one-dimensional disordered potential.
For the sake of simplicity, we shall restrict our consideration to the
propagation of one single atom, and thereby discard collective processes 
arising within Bose-Einstein condensates due to atom-atom interactions or 
superradiance.
We shall, moreover, assume that the atom will emit photons into one
single mode of the cavities only.
Such an emission will then give rise to a recoil that is mainly 
perpendicular to the direction of propagation of the atom and does 
therefore not dramatically enhance its longitudinal kinetic energy.
We neglect effects of transverse excitations within the waveguide due to
this recoil and assume that neither the effective waveguide confinement 
nor the disordered potential are affected by temporary populations of the
excited electronic state of the atom.

The dynamics of the atom is modeled via a one-dimensional master equation 
for its density matrix $\op{\rho}(t)$, which can be unraveled using the
Monte-Carlo wavefunction technique \cite{Unravel3,Unravel4}.
This master equation accounts both for the coherent motion within the
disordered potential and the incoherent scattering of photons \cite{Schloss}.
We shall, in Section \ref{sec:Wavepacketexpansion}, first account on the
expansion and localization dynamics of a single atom in a one-dimensional 
disordered potential in the absence of any decoherence mechanism.
In Section \ref{sec:MasterEquation}, we outline the Monte-Carlo wavefunction 
approach that is used to integrate the master equation for the special case 
of an atom that propagates in a homogeneous, disorder-free waveguide.
Decoherence and disorder are finally put together in Section \ref{sec:disdis},
in which we discuss the expansion of an atomic wave packet in the presence of 
disorder and spontaneous emission.
We show that even very rare position measurements of the atom give, on
average, rise to a gradual delocalization of the wave packet, and
we provide numerical evidence for superballistic expansion 
in the presence of strong emission rates.

\section{Wave packet expansion in disorder}
\label{sec:Wavepacketexpansion}

In this section, the expansion of an initially trapped wave packet in a weak 
one-dimensional disordered potential is discussed. 
For the sake of simplicity, we model the disorder by a Gaussian correlated 
random potential $V(x)$ defined along the $x$-axis, with the properties 
$\overline{V(x)} =0$ and
\begin{eqnarray}
\overline{V(x)V(x')} =U^2 \exp[-(x-x')^2/(2\sigma^2)]
\end{eqnarray}
for the mean spatial correlation function. 
Here, $U$ characterizes the typical size of the fluctuations of the potential,
and the correlation length $\sigma$ controls the average width of fluctuations.

\begin{figure}
\centering
\includegraphics[width=0.9\linewidth]{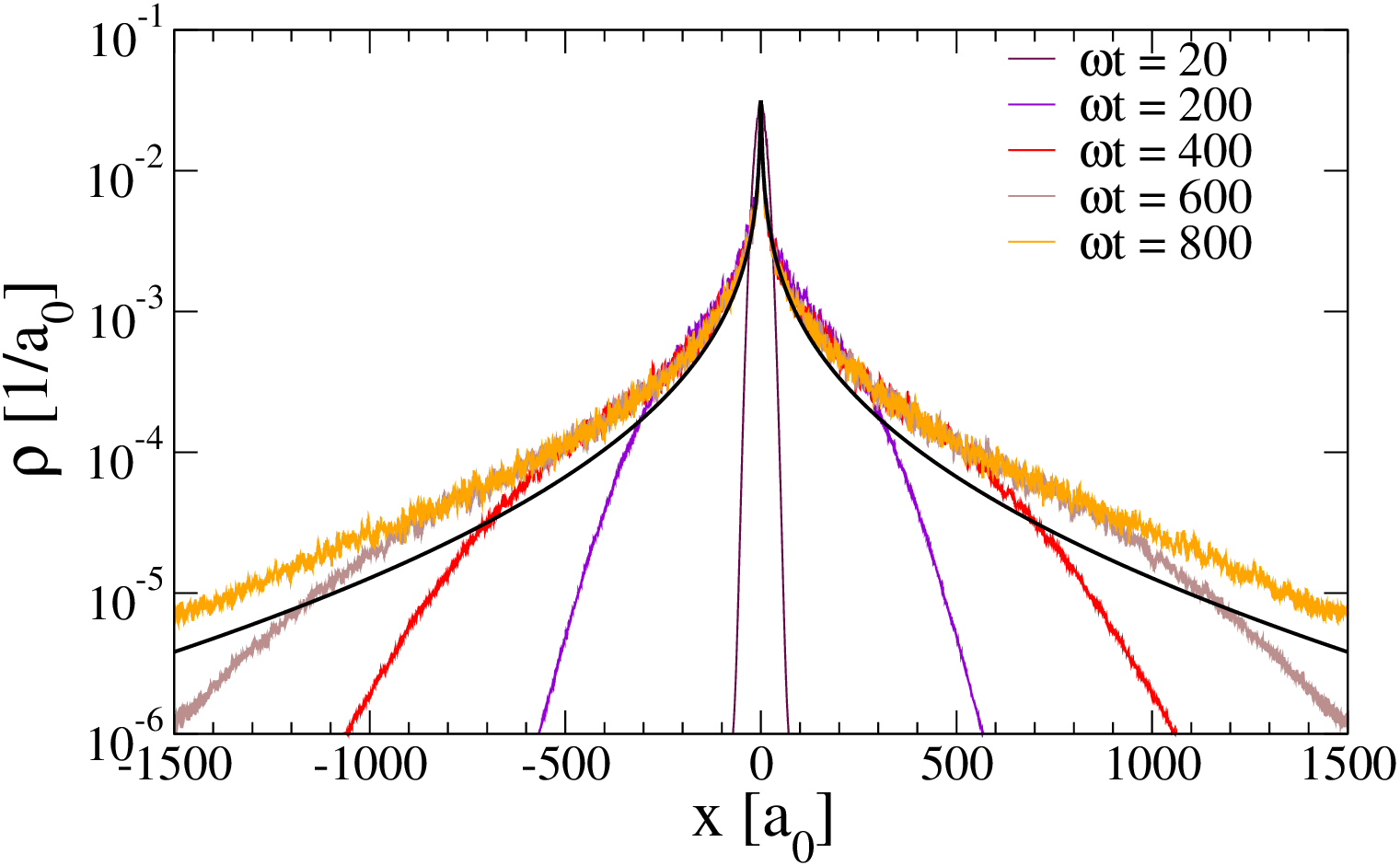}
\caption{(Color online)
Time evolution of the expansion process of a Gaussian wavefunction, 
averaged over 100 disorder realizations, for the strength
$U=0.1\,\hbar \omega$ and the correlation length $\sigma=0.2\,a_0$
of the disordered potential
(the mean initial kinetic energy of the wavefunction is 
$E=0.25\,\hbar \omega$, where $a_0$ and $\omega$ are the 
oscillator length and the frequency of the harmonic confinement 
potential, respectively).
Shown are the disorder-averaged probability densities $\rho(x)$ for
$\omega t = 20$, $200$, $400$, $600$, and $800$
(with the trap opening at $t=0$).
The solid line displays the analytical prediction (\ref{squaredecay})
which is found to be in good agreement with the numerical density 
distribution at large distances $|x| \gg a_0$, apart from a global
prefactor of the order of $2$.}
\label{timeloc}
\end{figure}

In Fig.~\ref{timeloc} we show the time evolution of the disorder-averaged 
spatial density of wave packets propagating in such disorder configurations.
These wave packets are initially prepared in the ground state of a 
harmonic trap with the oscillator length $a_{0}=\sqrt{\hbar/m\omega}$. 
After the trapping potential is switched off, the wave packet expands within
the disordered potential until it approaches, on average, a stationary profile.
The convergence to the average density distribution happens faster at the 
center than in the wings. 
This is a consequence of the quadratic growth of the localization length 
as a function of the wave vector, as described in Eq.~(\ref{Length}) below.

The final density profile shown in Fig.~\ref{timeloc} is fairly well 
reproduced by a theory as described, for example, 
in Ref.~\cite{Mueller}, which is based on the assum{p}tion that the 
asymptotic probability distribution is an
incoherent sum of individually localized plane waves with momentum $p$.
This consideration yields the spatial density
\begin{equation}
\rho_\mathrm{loc}(x) = \int \mathrm{d}p 
\frac{\rho_0(p)}{2\xi(p)}\exp[-|x|/\xi(p)]
\label{loc}
\end{equation}
where $\rho_0$ denotes the momentum density of the wave packet at the initial
time $t=0$.
The key ingredient for the evaluation of Eq.~(\ref{loc}) is the localization 
length $\xi(p)$ that can be calculated using diagrammatic theory \cite{Kuhn2} 
as $\xi(p)=2l_B(p)$, with $l_B$ the Boltzmann mean free path (see
Refs.~\cite{DisBec7,LSP1} for other approaches).
For the Gaussian correlated random potential under consideration, we obtain
\begin{equation}
\xi(p)= \frac{1}{\sqrt{2\pi}} \frac{\h^2p^2}{m^2U^2\sigma}
\exp[2(p\sigma/\hbar)^2]\,.
\label{Length}
\end{equation}
In the regime of short correlation lengths $\sigma \ll \hbar / p$, 
we can approximate $\exp[2(p\sigma/\hbar)^2] \simeq 1$ and the localization 
length depends only on the effective strength $U^2\sigma$ of the disorder.
Using
\begin{equation}
\rho_0(p) = \frac{a_0}{\sqrt{\pi} \hbar} \exp[ - (a_0 p / \hbar)^2 ]
\label{rho0}
\end{equation}
and introducing the characteristic localization length scale of the 
wave packet as $\xi_0 \equiv \xi(\hbar/a_0)$, 
we then obtain the prediction
\begin{equation}
\rho_\mathrm{loc}(x) = \frac{1}{2\sqrt{\xi_0|x|}} 
\exp\left( - 2 \sqrt{|x|/\xi_0} \right)
\label{squaredecay}
\end{equation} 
for the localized density.
As shown in Fig.~\ref{timeloc}, this approximate expression is,
apart from a global prefactor, in good agreement with the numerically 
computed mean density at the final time $t=800/\omega$.

In the above numerical simulations, we effectively assumed that the atomic
cloud is prepared in a clean harmonic trap in absence of any disorder.
At $t=0$ the trapping potential is suddenly switched off and the disorder
is ramped on at the same time.
The initial state is then a perfect Gaussian wavefunction 
[see Eq.~(\ref{rho0})] which expands within the disordered potential.
This procedure is, in general, not precisely in accordance with expansion 
experiments on Anderson localization such as Ref.~\cite{Nature1} in which
the disordered potential is already present {during the formation} of the 
Bose-Einstein condensate in the harmonic trap.
The initial state of the atomic wavefunction is, in that case, given by the 
ground state of an effective trapping potential that consists of a harmonic 
confinement modulated by the disorder.
A numerical comparison of these two expansion scenarios, however,
displays no significant difference in the asymptotic density profile
for the case of weak disordered potentials with $U \simeq 0.1\, \hbar\omega$
and $\sigma = 0.2\, a_0$.

\begin{figure}[t]
\centering
\includegraphics[width=0.8\linewidth]{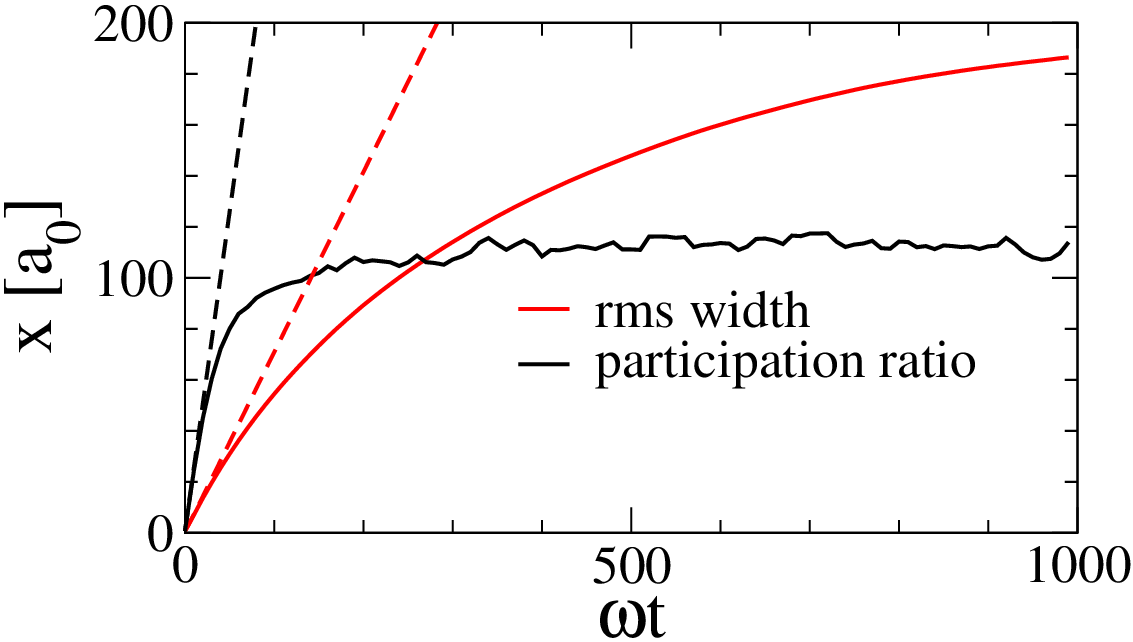}
\caption{(Color online) Root mean square (rms) width
$\Delta x = \sqrt{\langle x^2\rangle - \langle x \rangle^2}$ (red line)
and participation ratio $Pr(t)$ (black line) 
as a function of the evolution time $t$,
showing the expansion and localization of a wave packet
for the disorder strength $U=0.15\,\hbar \omega$
and the correlation length $\sigma=0.2\, a_0$.
The dashed lines show, for comparison, the rms width and the 
participation ratio of a free wave packet that expands in the absence 
of disorder.}
\label{part}
\end{figure}

A convenient numerical observable for measuring localization is the 
participation ratio \cite{Kramer1993} which for a wave packet 
with the density $\rho(x,t)$ is defined by
\begin{equation}
\mathrm{Pr}(t) = \left( \int_{-\infty}^{\infty}dx 
[\rho(x,t)]^2 \right)^{-1}\, . \label{eq:part}
\end{equation}
In practice, $\mathrm{Pr}(t)$ represents a measure for the spatial extent of
the wave packet, yielding large values for rather extended distributions 
$\rho(x,t)$ and going to zero for strongly peaked wavefunctions.
It therefore exhibits a similar behavior to the spatial root mean square 
(rms) width 
$\Delta x = \sqrt{\langle x^2 \rangle - \langle x \rangle^2}$ of the
wave packet.
This latter quantity is, however, rather sensitive to the evolution of the
(experimentally inaccessible) wings of the wave packet.
This is shown in Fig.~\ref{part} where we display the time dependence of the 
disorder-averaged rms width and participation ratio.
While the rms width continuously increases with time, due to the 
long-time dynamics in the wings of the averaged density distribution 
(see Fig.~\ref{timeloc}), the participation ratio, which is much less 
sensitive to the behavi{o}r of the wings, saturates at a finite length scale.
This length scale can be used in order to define an effective localization 
length $L_{\mathrm{wp}}$ of the wave packet.

\begin{figure}[t]
 \centering
\includegraphics[width=\linewidth]{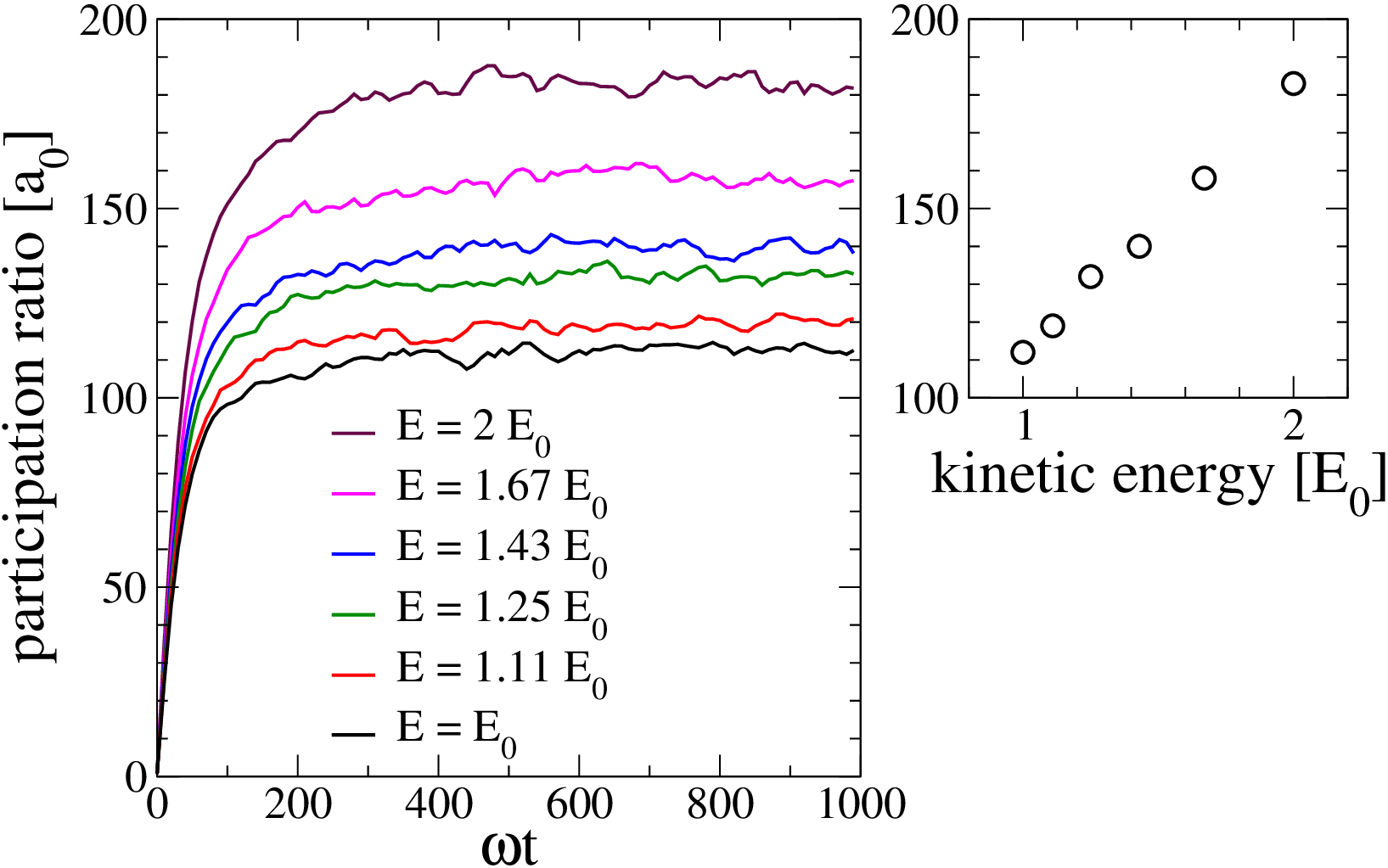}
\caption{(Color online)
Left panel: participation ratio as a function of time for 
different initial kinetic energies $E$, with $E_0=0.25\,\hbar \omega$.
The disordered potential is characterized by the parameters 
$U=0.15\,\hbar \omega$ and $\sigma=0.2\, a_0$.
The right panel shows the average of the participation
ratio within $400 < \omega t < 1000$ as a function of the energy $E$.
This average defines the asymptotic localization length 
$L_{\mathrm{wp}}$ of the wave packet.
$L_{\mathrm{wp}}$ is found to increase approximately linearly with the
initial kinetic energy.
\label{energydep1}}
\end{figure}

Fig.~\ref{energydep1} shows the time evolution of the participation ratio 
for different initial kinetic energies $E$, resulting from different 
confinement frequencies of the initial trapping potential.
A linear increase of the participation ratio with $E$, corresponding to a 
linear increase of the effective localization length $L_{\mathrm{wp}}$ of the 
wave packet, is found for $E_0 < E < 2 E_0$.

\section{Master equation dynamics}
\label{sec:MasterEquation}

To account for spontaneous emissions of photons into the cavities, 
we model the dynamics of the atom via a one-dimensional master 
equation for its density matrix $\op{\rho}(t)$, including coherent 
interactions with a disordered potential and the incoherent scattering 
of light \cite{Schloss}.
This master equation is given by
\begin{equation}
\frac{d}{dt}\rho(t) = - {\frac{i}{\hbar}} 
[\op{H},\op{\rho}(t)] + \gamma_{\mathrm{eff}} \int_{-k}^{k}\frac{\mathrm{d}q}{2k} \left( \op{C_q}\op{\rho}(t)\opdag{C_q} - \op{\rho}(t) \right)\,.
\label{MasterEq}
\end{equation}
Here, $\op{H}=\frac{\op{p}^2}{2m} + V(\op{x})$ describes the Hamiltonian
for a particle that propagates in the disordered potential.
$\op{C}_q= e^{-iq\op{x}}$ is the decay or jump operator representing one 
spontaneous emission event, which exerts a recoil on the atom with 
longitudinal momentum $\hbar q$ which is assumed to be equidistributed 
between $-\hbar k$ and $+\hbar k$.
This model considers off-resonant inelastic scattering, in which a
laser couples the electronic ground state to an excited state from
which spontaneous emission back to the ground state can occur.
It assumes a low spontaneous decay rate $\gamma$ as compared to the 
detuning $\delta$ of the laser with respect to the intra-atomic transition 
frequency, and a low Rabi frequency $\nu$ for laser-induced
transitions between the ground state and the excited state as compared to the 
spontaneous decay rate $\gamma$, i.e.\ we assume
$\nu \ll \gamma \ll \delta$.
We then obtain $\gamma_{\mathrm{eff}} = \gamma \nu^2/(\gamma^2+4\delta^2)$
as the effective decay rate that enters the master equation (\ref{MasterEq})
\cite{Adiabatic1,Adiabatic2}.

To solve the time evolution generated by the master equation, 
we employ the Monte Carlo wave function method \cite{Unravel3,Unravel4}. 
Here, the evolution of the density matrix is decomposed into the 
non-unitary evolution of a large number $N = 100$ of wave functions. 
A single trajectory $|\psi\rangle_i,\, i=1...N$ evolves according to
$i\hbar \del_t|\psi\rangle_i = \op{H}_\mathrm{eff}|\psi\rangle_i $ with 
$\op{H}_\mathrm{eff} \equiv \op{H}-i\gamma_\mathrm{eff}/2$,
until the exponentially decaying norm 
$||\psi\rangle_i|^2 = e^{-\gamma_\mathrm{eff}t}$ equals a random number 
chosen between $0$ and $1$. 
At this point, a jump operator $\op{C}_q$ acts on the Monte Carlo wave 
function: $|\psi(t+\delta t)\rangle_i =\op{C}_q |\psi(t)\rangle_i$. 
This jump operator is determined by randomly choosing $q$ from the 
interval between $-k$ and $k$.

To relate this light scattering process to the position measurement under
consideration, we note that the Lindblad master equation is invariant under 
unitary transformations on the set of decay operators. 
Indeed, it was shown in Ref.~\cite{Murray} that the Fourier transformation 
$\int_{-1}^{1}du \, \exp(iuk\nu\lambda/2) \,\op{C}_u$ with integers 
$\nu\, \in \mathbb{Z}$ allows one to switch to decay operators
\begin{equation}
\op{C}_{\nu}=\sqrt{2}\frac{\sin(k\hat{x}-\frac{\nu}{2})}{k\hat{x}-\frac{\nu}{2}}
\end{equation}
In this picture, the application of the decay operator induces a 
localization of the wavefunction within a spatial region whose extent is 
of the order of $k^{-1}$.
For the sake of simplicity, we assume that these decay operators exactly
correspond to the longitudinal structure of the cavity modes into which the
atom may emit the photon.
The spatial period of the array of cavities is then given by 
$\lambda = 2 \pi / k$.

\begin{figure}[t]
\centering
\includegraphics[width=0.6\linewidth]{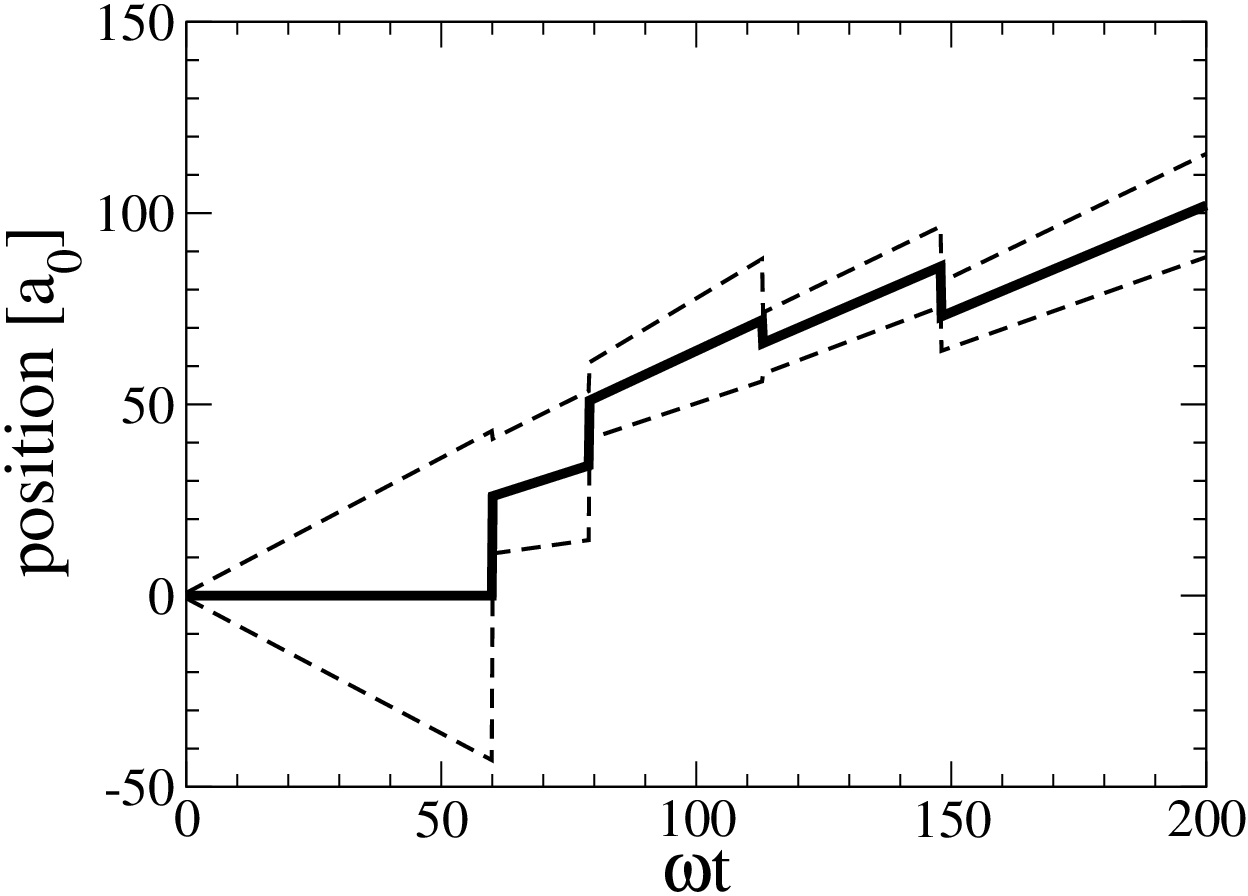}
\caption{
Center of mass position $\langle x \rangle$ (solid line) 
for a single trajectory of an atom that subject to light scattering
with $k=0.07/a_0$.
The dashed lines indicate the root mean square (rms) width
$\Delta x = \sqrt{\langle x^2 \rangle - \langle x \rangle^2}$,
i.e.\ they correspond to the lines $\langle x \rangle \pm \Delta x$.
Photon emissions occur at $\omega t=60$, $79$, $113$, and $148$.}
\label{freetrajectory}
\end{figure}

In Fig.~\ref{freetrajectory}, we show the time evolutions of the expectation 
value of the position $\langle x \rangle$ and its rms width
$\Delta x = \sqrt{\langle x^2 \rangle - \langle x \rangle^2}$ 
for a single quantum trajectory in free space, $V(x)=0$.
In this particular trajectory, the first spontaneous emission took place in 
one of the two wings of the wave packet, which is mainly constituted by
plane-wave components with high momenta.
The subsequent localization process projects the wavefunction on those 
high-momentum components, which gives rise to a permanent drift. 
The rms width, however, remains small during this evolution, which is due to
the fact that the atom emits photons at a rate that is faster than the 
inverse dispersion time of the wave packet.
The rms width would freeze for sufficiently high emission rates, which is 
reminiscent of the quantum Zeno effect. 
 
It is of great advantage to work in a regime where $k$ is small compared 
to fluctuations of the density matrix in momentum space $\dm{}{p,}{p',}{t}$. 
To study the momentum density distribution, we can then approximate the 
integrand of Eq.~(\ref{MasterEq}) by its Taylor expansion to first order, as
done in Ref.~\cite{Joos1}. 
Taking the integral over $k$ leads to the diffusion equation
\begin{equation}
\partial_t\dm{}{p,}{}{t}=\frac{1}{6}\gamma_{\mathrm{eff}} k^2\partial_p^2\dm{}{p,}{}{t}\,.
\label{prox}
\end{equation}
for the diagonal elements of the density matrix, with the effective 
diffusion coefficient $D=\gamma_\mathrm{eff} k^2/6$.
Hence, the wave packet will undergo diffusive spreading in momentum space. 
Noting that the variance of the momentum distribution is nothing but the 
kinetic energy, we obtain
\begin{equation}
\langle \op{T} \rangle = \mathrm{Tr}\{\frac{\op{p}^2}{2m}\op{\rho}(t)\}=
E_0 + \frac{ \hbar^2 k^2}{6m}\gamma_\mathrm{eff} t
\label{energygain}
\end{equation}
for the growth of the mean kinetic energy of the wave packet.

\section{Dissipative expansion in disorder}
\label{sec:disdis}

\begin{figure}
\centering
\includegraphics[width=\linewidth]{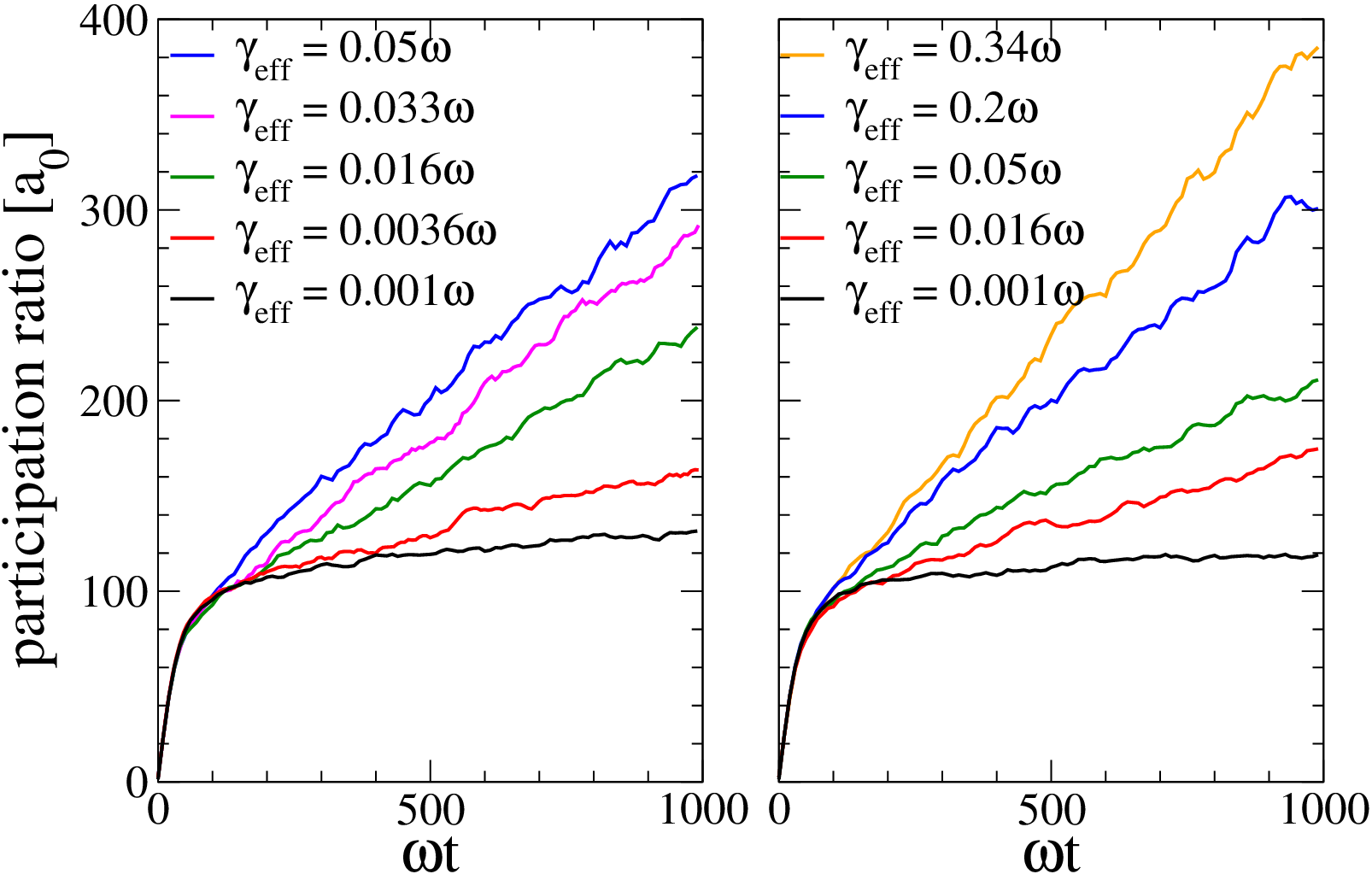}
\caption{(Color online) 
  Master equation dynamics of the participation ratio for different 
effective emission rates $\gamma_{\mathrm{eff}}$. 
The longitudinal momentum of the emitted photon is
$k=0.07/a_0$ in the left panel and $k=0.035/a_0$ in the right panel.
The linear growth of the participation ratio for $\omega t \gtrsim 100$
reflects the growth of the mean kinetic energy due to photon scatttering 
events, which is found to increase from $E = 0.25\,\hbar\omega$ at $t=0$ to
$E \simeq 0.5\,\hbar\omega$ at $\omega t = 1000$ for
$k=0.07/a_0$ and $\gamma_{\mathrm{eff}} = 0.05 \omega$
(blue curve in the left panel).
The disordered potential is characterized by the strength 
$U=0.15\,\hbar\omega$ and the correlation length 
$\sigma=0.2\,a_0$.}
\label{masterdynamics1}
\end{figure}

Having introduced the necessary tools, we now study wave packet expansion 
in the presence of disorder and dissipation. 
In Fig.~\ref{masterdynamics1} we plot the participation ratio as a function 
of time for different effective emission rates $\gamma_{\mathrm{eff}}$. 
In accordance with the sketch shown in Fig.~\ref{setup}, we
have chosen the photon wavelength to be very long compared to
the initial extension $a_0$ of the wave packet.
The amount of kinetic energy given to the wave packet at each emission 
event is thereby rather reduced.

The most important observation is a delocalization of the wave packet at 
any emission rate.
Instead of saturating to a stationary value, the participation
ratio linearly increases with time after the typical time scale that is
needed for developing an Anderson-localized density profile in the
absence of spontaneous emission.
Quantitatively, this linear growth is very different from a ballistic 
expansion in free space, which takes place with much faster expansion 
velocities (see the dashed lines in Fig.~\ref{part}).
It is also different from simple diffusion which one would naively 
expect to prevail for a quantum particle that propagates within a
disordered potential in the presence of a decoherence mechanism.
We attribute this difference to the fact that the spontaneous emission of 
a photon gives rise to a recoil of the atom and thereby increases its energy.
Hence, the effective diffusion constant should also gradually increase with 
time.

It is, in this context, interesting to note that the expansion velocity
${d\mathrm{Pr}/dt}$ depends only on the product of the effective rate of
emission $\gamma_{\mathrm{eff}}$ and the recoil energy $\hbar^2 k^2/(2m)$.
This can be seen by comparing the two blue lines in the two panels of 
Fig.~\ref{masterdynamics1} showing expanding participation rates for 
$k=0.07/a_0$ and $\gamma_{\mathrm{eff}}=0.05\omega$ (left panel) as well
as for $k = 0.035/a_0$ and $\gamma_{\mathrm{eff}} =0.2\omega$.
There appears, furthermore, no change in the behavior when we tune the rate 
of emissions across the scale $1/T_\mathrm{loc}$, with $T_\mathrm{loc}$ the 
time at which the unperturbed evolution shows localization. 

It is tempting to relate the linear increase of the participation rate with
time to the combination of a linear growth of the kinetic energy due to 
spontaneous emission with the approximately linear scaling of the 
wave packet's localization length with its mean kinetic energy in the absence 
of spontaneous emission, as shown in Fig.~\ref{energydep1}.
This reasoning essentially assumes that in between two subsequent spontaneous
emission events the wave packet has enough time to approach its asymptotic
stationary profile within the disordered potential.
Extracting from Fig.~\ref{energydep1} the approximate scaling 
$Pr/a_0 \sim 100 E/E_0$ and using 
$dE/dt = \hbar^2 k^2 \gamma_{\rm eff} / (6m)$ for the growth rate
of the energy according to Eq.~(\ref{energygain}), we obtain the prediction
\begin{equation}
\frac{d\mathrm{Pr}}{dt} 
\simeq 100\,\frac{a_0}{E_0}\, \frac{dE}{dt} \simeq
400\, \frac{\gamma_{\rm eff}\hbar^2 k^2}{6m}\, \frac{a_0}{\hbar \omega}
\label{linexp}
\end{equation}
for the expansion velocity ${d\mathrm{Pr}/dt}$ of the participation rate,
using $E_0 = 0.25 \hbar \omega$.

\begin{figure}[t]
 \centering
\includegraphics[width=\linewidth]{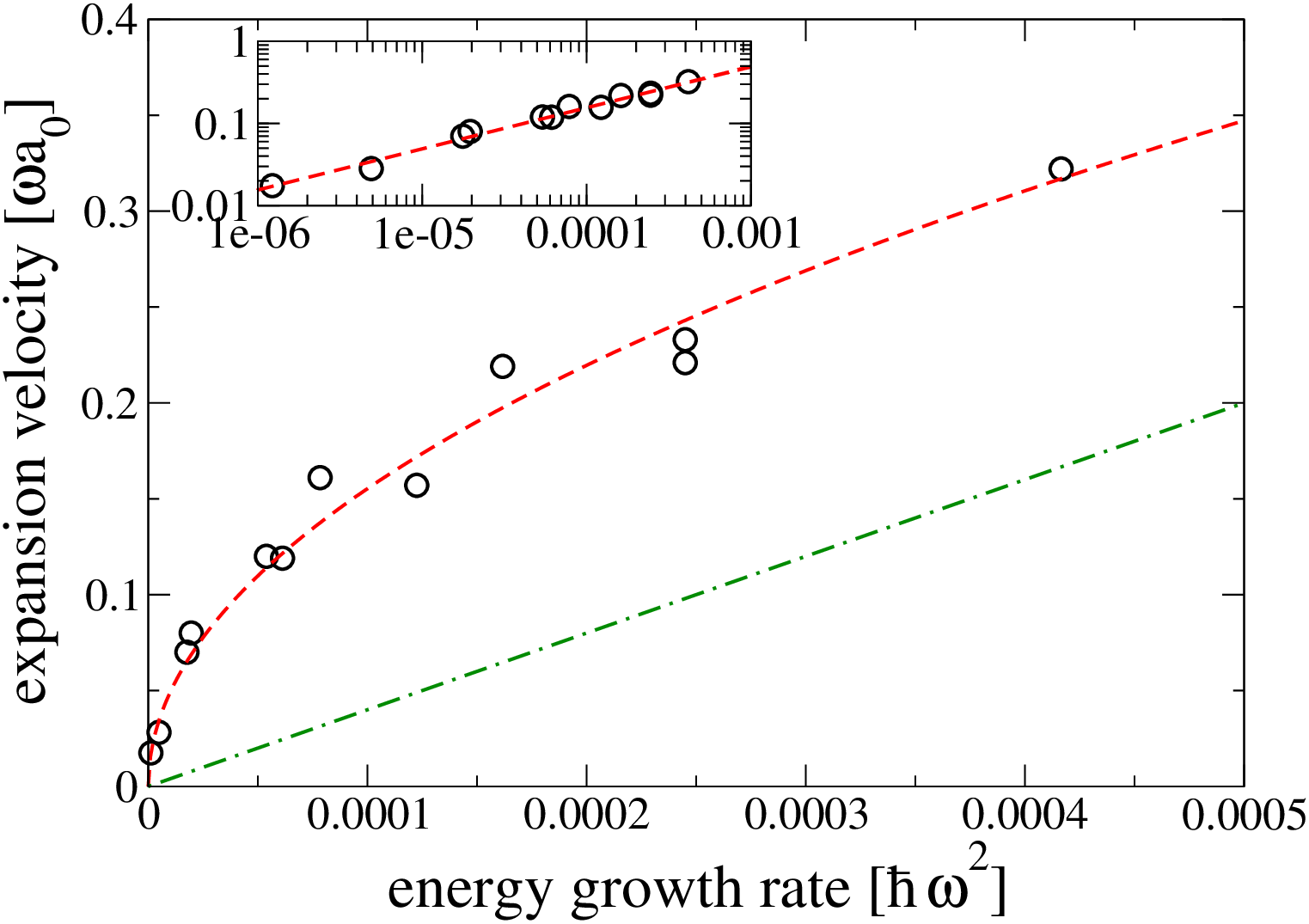}
\caption{(Color online) 
Expansion velocity ${d\mathrm{Pr}/dt}$ of the participation ratio
as a function of the energy growth rate
${dE/dt} = \gamma_{\mathrm{eff}} \hbar^2 k^2/(6m)$. 
The data are extracted from Fig.~\ref{masterdynamics1}, as well as from other
calculations using different parameters for $\gamma_{\mathrm{eff}}$ and $k$,
through linear regression of the participation ratio within 
$100<\omega t<1000$.
As confirmed in the log-log plot shown in the inset, 
${d\mathrm{Pr}/dt}$ scales as the square root of ${dE/dt}$: 
we have ${d\mathrm{Pr}/dt} \simeq \alpha {(dE/dt)^{1/2}a_0 /\hbar^{1/2}}$
with the fitted proportionality constant $\alpha \simeq 15.5$, as indicated
by the dashed line.
The dash-dotted straight line in the main panel represents the prediction of
Eq.~(\ref{linexp}).
\label{ExpansionRate}}
\end{figure}

Figure \ref{ExpansionRate} shows, however, that this expansion velocity 
increases more strongly with the rate of increase of the kinetic energy
than predicted by Eq.~(\ref{linexp}).
As a matter of fact, ${d\mathrm{Pr}/dt}$ is found to scale as a square 
root of ${dE/dt}$ in the parameter regime in which we carried out our 
numerical investigations.
One may attribute this behavior to the fact that the above 
reasoning rather applies to an \emph{individual} quantum trajectory in the 
spirit of Fig.~\ref{freetrajectory}.
The energy of the wavepacket corresponding to each individual
trajectory increases linearly and its participation ratio increases on 
average as described by Eq.~(\ref{linexp}). 
However, while different trajectories describe similar narrow wavepackets, 
each wavepacket will be centered around a different point in space. 
Thus the full (incoherent) density will be spreading faster
over a larger region than a single wavepacket (as is obvious from 
Fig.~\ref{freetrajectory} for the case of disorder-free propagation).
This effect is obviously not accounted for in the considerations leading to
Eq.~(\ref{linexp}).

\begin{figure}[t]
\centering
\includegraphics[width=0.7\linewidth]{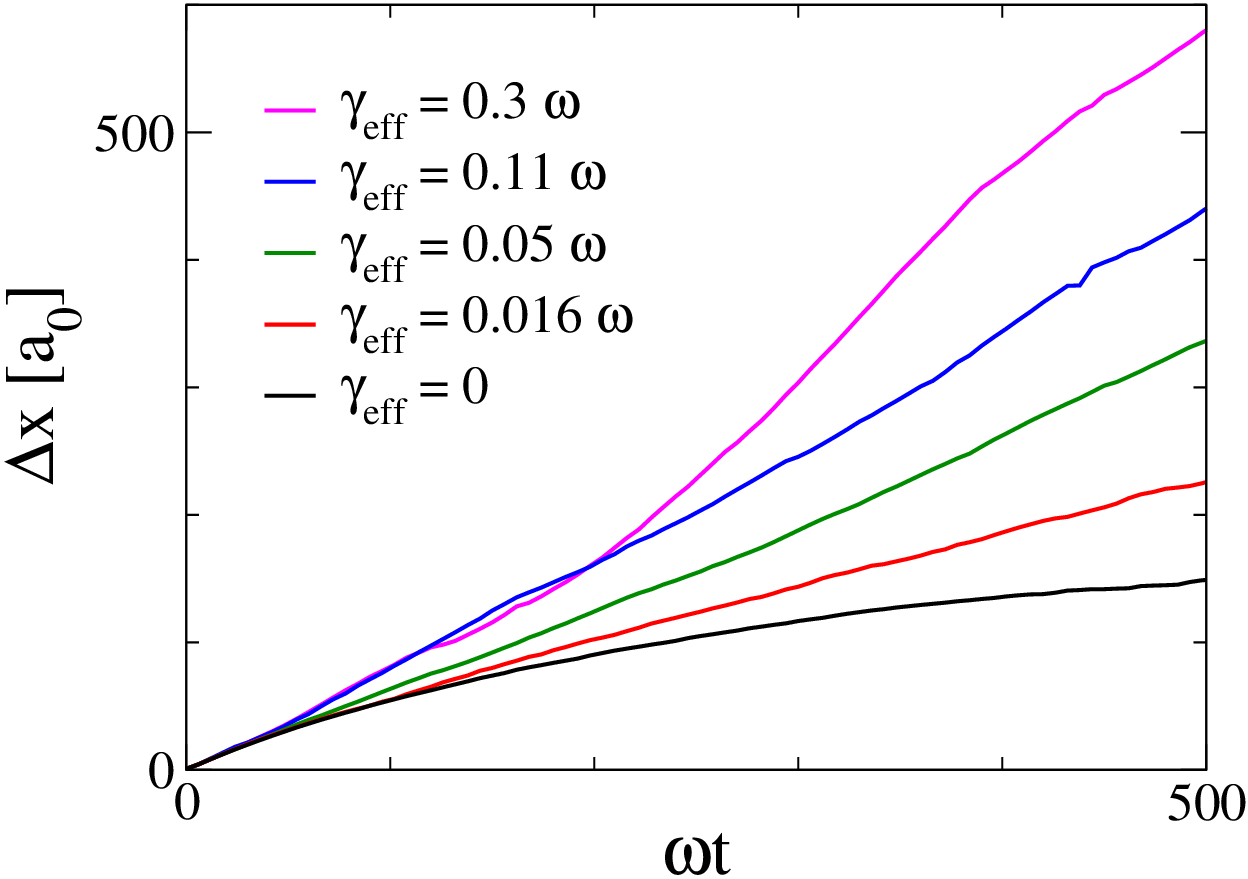}
\caption{(Color online)
Master equation dynamics of the rms width $\Delta x$ at strong recoil 
$k = 0.35/a_0$ for different effective emission rates $\gamma_{\mathrm{eff}}$.
{The disordered potential is characterized by the parameters 
$U=0.15\,\hbar \omega$ and $\sigma=0.2\,a_0$.}
We note that the growth of $\Delta x$ for 
$\gamma_{\mathrm{eff}} = 0.05$ $\omega$ (green curve) 
coincides with the one of a free ballistic expansion,
while superballistic expansion is encountered for larger emission rates.}
\label{disk5}
\end{figure}

Let us finally investigate the regime of strong dissipation for which it is 
expected that the expansion becomes independent of the disordered potential.
In Fig.~\ref{disk5} the momentum recoil is set to $k=0.35/a_0$, leading to
a regime in which the rms width turns out to serve as an accurate measure of 
expansion. 
In this case, the recoil induced by the emitted photons may drive the 
system beyond the free ballistic growth into a superballistic regime, 
which is the valid limit of a driven expansion in free space \cite{Joos1}. 
As can be seen in Fig.~\ref{disk5}, this superballistic regime sets in
beyond $\gamma_{\mathrm{eff}} = {0.05\,\omega}$ for $k=0.35/a_0$. \\

\section{Conclusion}
\label{sec:summary}

In summary, we have shown that even a very weak rate of photon
scattering gives rise to a breakdown of Anderson localization of an atom
that propagates in a one-dimensional disordered potential.
This breakdown is most conveniently quantified in terms of the 
disorder-averaged participation ratio of the atomic density, which 
represents a measure for the spatial width of the atomic wave packet.
While this participation ratio saturates, within a characteristic time
scale, to a finite value in the case of a perfectly coherent expansion 
process within the disordered potential, it is found to linearly grow with 
time beyond that time scale in the presence of spontaneous photon scattering.

This growth behavior imposes strong limits for the observability of
Anderson localization in the presence of a weak position measurement
of the atom.
However, an experimental realization of a ``Heisenberg microscope'' for 
cold atoms according to the scheme displayed in Fig.~\ref{setup} might 
nevertheless be of interest as it allows one to study in more detail the 
interplay of disorder and measurement-induced delocalization phenomena 
not only for a single atom, but also (and this more naturally) for a 
Bose-Einstein condensate in which the atoms interact with each other.
For this purpose, an integrated setup on atom chips appears as the most 
convenient realization of such a Heisenberg microscope for atomic gases.

Finally, we expect similar findings in the presence of other mechanisms 
that can behave as a position measurement of the propagating atom.
Such mechanisms include noise on the lattice beams as well as collisions with
background gas atoms, to mention two examples.
Undesired effects of this type are therefore also expected to induce a 
delocalization of the atom in the disorder potential.

\acknowledgements

The authors would like to thank K.~Hornberger, C.~A.~M\"uller, 
and B.~M.~Peden for useful discussions.
M.~H.~acknowledges support from the National Science Foundation.

\end{document}